\documentclass[twocolumn,showpacs,amssymb,prd]{revtex4}

\usepackage{amsmath}
\usepackage{latexsym}
\usepackage{graphicx}
\usepackage{epstopdf}
\usepackage{verbatim}

\begin{document}


\title{Spin coefficients and gauge fixing in the Newman-Penrose formalism} 


\author{Andrea Nerozzi$\footnote{Electronic address: andrea.nerozzi@ist.utl.pt}
$}
\vskip 3mm
\affiliation{Centro Multidisciplinar de Astrof\'isica - CENTRA, Departamento de F\'isica, Instituto
Superior T\'ecnico - IST, Universidade de Lisboa - UL,
Avenida Rovisco Pais 1, 1049-001 Lisboa, Portugal}
%
%


\date{\today}


\begin{abstract}

Since its introduction in 1962, the Newman-Penrose formalism has been widely used in analytical and 
numerical studies of Einstein's equations, like for example for the Teukolsky master equation, 
or as a powerful wave extraction tool in numerical relativity. 
Despite the many applications, Einstein's equations in the Newman-Penrose formalism appear complicated
and not easily applicable to general studies of spacetimes, mainly because physical and gauge degrees
of freedom are mixed in a nontrivial way. 
In this paper we approach the whole formalism with the goal of expressing the spin coefficients
as functions of tetrad invariants once a particular tetrad is chosen.   
We show that it is possible to do so, and give for the first time a general recipe for the task,
as well as an indication of the quantities and identities that are required. 
\end{abstract}




\maketitle

In 1962 Newman and Penrose \cite{Newman:1962el} presented a new tetrad approach
to Einstein's equations based on null tetrad vectors. The relevant equations, namely the Bianchi
and Ricci identities were determined, together with an alternative demonstration
of the Goldberg-Sachs \cite{Goldberg:2009hv} theorem and the study of the asymptotic behaviour of the
Riemann tensor for asymptotically flat spacetimes in vacuum. 

Since its introduction the Newman-Penrose (NP) formalism proved to be a powerful
approach to Einstein's equations studied in several areas of general relativity. In 1973
Teukolsky \cite{Teukolsky:1973kf} formulated his famous master equation based on the NP formalism
giving decoupled perturbation equations for two Weyl scalars $\Psi_0$ and $\Psi_4$. This strengthened the idea of these scalar
fields being associated with the gravitational waves degrees of freedom, respectively ingoing
and outgoing, a result that had been already
anticipated by Newman and Penrose in their seminal paper.

With the advent of numerical relativity the NP formalism found another important application:
a tool for gravitational wave extraction in numerical simulations (for an exhaustive review
on wave extraction methods see \cite{Rezzolla:2016}).
Given its tight association to the gravitational wave degrees of freedom and its coordinate invariant
properties, the calculation of $\Psi_4$ in a numerical grid seemed to be the most natural candidate
for a rigorous wave extraction methodology. However, the freedom
in the choice of tetrads constitutes a possible source of undesired gauge effects, which
led to a series of papers on the topic aimed at finding the most rigorous approach. The main motivation underlying these works 
was to define a gauge invariant quantity associated with gravitational waves. 
Beetle and Burko \cite{Beetle:2002gt} published a paper
in 2002 identifying a radiation scalar with interesting properties for
wave extraction, following a previous work by Baker and Campanelli \cite{2000PhRvD..62l7501B} which proposed
that a certain function of curvature invariants, the speciality index, could be studied as an invariant
measure of distortions of spacetimes. These works were soon followed by a series of papers in the field aiming
to identify
an optimal tetrad in which to calculate $\Psi_4$ (or $\Psi_0$ for ingoing waves). 
This special choice was named the ``quasi-Kinnersley" tetrad \cite{Beetle:2004wu, Nerozzi:2004wv, 2006PhRvD..73d4020N, 2007PhRvD..75j4002N} because of its natural property of converging 
to the Kinnersley tetrad \cite{1969JMP....10.1195K} in the single black hole limit. This tetrad
was found to be part of a particular set of tetrads that were dubbed ``transverse"
tetrads, namely those in which $\Psi_1=\Psi_3=0$. Incidentally this definition corresponds
to the ``canonical" frame previously introduced by Edgar, Brans and Bonanos \cite{Brans:1977jd, 
Bonanos:1996uy, Bonanos:1991dl}.

The concept of a quasi-Kinnersley tetrad has been implemented in numerical simulations \cite{2006PhRvD..73f4005C}
and as a tool to invariantly characterise numerically evolved spacetimes \cite{2012PhRvD..86h4020Z}. However, 
its definition suffers from the indetermination of the spin/boost
parameter. The reason is simple: the Kinnersley tetrad for a Kerr black hole
was derived by imposing a specific condition on one spin coefficient, namely
$\epsilon=0$. 
In order to enforce this condition in a numerical spacetime, i.e., for a generic
Petrov type I spacetime, one needs a well-defined expression for all the spin coefficients
in transverse frames.
Some more recent
works \cite{Nerozzi:2008ng} gave a first attempt to solve this problem, however limited
to the case of Petrov type D spacetimes. The present paper solves the problem for a general
Petrov type I spacetime and gives a recipe to express all of the spin coefficients as functions
of tetrad invariants when transverse tetrads are considered.  

The possible applications of the results found in this paper go well beyond the problem of wave
extraction in numerical relativity. For example, it can give new insights for numerical studies of Einstein's
equations using tetrad approaches, for which there is already extensive literature, see e.g.
\cite{Estabrook:323016, 2011PhRvD..83j4045B, 2005PhRvD..72l4014B, 2003PhRvD..67h4017B}, as the
problem of gauge fixing within these approaches has not been faced in detail before. 
The successes of numerical relativity \cite{Pretorius:2005wl, Campanelli:2006gw, Baker:2006ww}  together with the recent exciting direct detection 
of gravitational waves \cite{Abbott:2016ki}, operated by Laser Interferometer Gravitational Wave Observatory, now motivates the study of new and 
more refined methodologies to obtain accurate gravitational wave templates, and tetrad approaches 
are certainly among those. Moreover, it can provide new ideas for solving open problems
in the generalisation of Einstein's equations to higher dimensions  
\cite{Coley::2004, 2012CQGra..29t5002O, 2013CQGra..30a3001O, 2007CQGra..24.1657O},
like the study of perturbations \'a la Teukolsky
\cite{2012PhRvD..85h4021G, 2012CQGra..29e5008G}.

The work presented here is to be 
considered the first of two steps aimed at expressing
all of the relevant quantities in the NP formalism as functions of tetrad invariants, i.e., quantities
that are not affected by any tetrad transformations and can be calculated in
any coordinate system, making them appealing for numerical calculations. The two curvature
invariants $I$ and $J$ are obvious examples of tetrad invariant quantities. If all of the gauge
degrees of freedom are removed from a tetrad formalism, all of the
remaining relevant quantities must be functions of tetrad invariants. 
Removing the gauge freedom in the NP formalism leads to 
the main result of this paper given by Eq.~(\ref{eqn:finalABC}) 
where the spin coefficients are obtained as functions of the curvature invariants
$\nabla_a I$ and $\nabla_a J$ plus
an additional tetrad invariant vector $\mathcal{S}_a$. 
The second step of our work will be presented in a follow-up paper and will give a more
rigorous 
characterisation of the vector $S_a$ by studying in detail Eq.~(\ref{eqn:bianchityperelforD}) which is key to Eq.~(\ref{eqn:finalABC}).
The applications to numerical relativity and to the problem of wave extraction will be discussed in the conclusions.

The paper is organized as follows: In Sec. \ref{sec:NPtransverse} the
NP formalism in transverse frames is
presented. 
It is shown that the Bianchi identities can be written in a compact way as was already found by Bonanos \cite{Bonanos:1996uy}. The Bianchi identities are however not enough to express
all the spin coefficients as functions of tetrad invariants. In order
to find the missing relations, 
in Sec. \ref{sec:NPselfdual} an approach to the NP formalism based on self-dual forms is
introduced. In Sec. \ref{sec:bianchiselfdual} the curvature will be analysed within the
self-dual form approach, in particular introducing the Laplacian of the self-dual Weyl tensor.
In Secs. \ref{sec:linkcurvselfdual} and \ref{sec:tetradinvariantchar} it will be shown that
the information on the divergences of the Weyl tensor and its Laplacian give a well-posed
system to express all the spin coefficients as functions of tetrad invariants. 
The calculation will be then performed in 
Sec. \ref{sec:dertetradindependent} where the final expression for the spin coefficients
will be given. Finally the Petrov type D limit is presented in Sec. \ref{sec:kerrexample} to prove the consistency of this new approach.

Throughout the paper a four-dimensional Lorentzian manifold is considered where tensor
components are labeled with latin indices, and where $\nabla_a$ is the standard
covariant derivative associated to the Levi-Civita connection.

\section{The NP formalism in transverse tetrads}
\label{sec:NPtransverse}

\subsection{Weyl scalars and curvature invariants}
\label{eqn:weylscalarscurvatureinv}

The relevant variables in the NP formalism are the
Weyl scalars representing the curvature and the
spin coefficients representing the connection. Weyl scalars are obtained by contracting the Weyl
tensor along different combinations of the null tetrad vectors $\ell^a$, $n^a$, $m^a$ and
$\bar{m}^a$, according to

\begin{subequations}
\label{eqn:weylscalars}
\begin{eqnarray}
\Psi_0 &=&-C_{abcd}\ell^am^b\ell^cm^d, \label{eqn:weylscalars0} \\
\Psi_1 &=&-C_{abcd}\ell^an^b\ell^cm^d, \label{eqn:weylscalars1} \\
\Psi_2 &=&-C_{abcd}\ell^am^b\bar{m}^cn^d, \label{eqn:weylscalars2} \\
\Psi_3 &=& -C_{abcd}\ell^an^b\bar{m}^cn^d, \label{eqn:weylscalars3} \\
\Psi_4 &=& -C_{abcd}n^a\bar{m}^bn^c\bar{m}^d. \label{eqn:weylscalars4} 
\end{eqnarray}
\end{subequations}

The tetrad vectors satisfy the contraction identities $\ell^a n_a = -1$ and 
$m^a\bar{m}_a=1$. The spin coefficients are twelve complex scalar quantities that
can be divided in the three groups $\{\rho,\mu,\tau,\pi\}$, $\{\lambda,\sigma,\nu,\kappa\}$
and $\{\epsilon,\gamma,\beta,\alpha\}$. It will be shown is Sec. \ref{sec:spincoeffsect} that each 
group can be expressed as projections along the tetrad vectors of a suitable vector. 
Each spin coefficient is associated with important features of the tetrad vectors (see
\cite{chandrasekhar1983mathematical} for details), so for example if the $\ell^a$ vector is geodesic, $\epsilon=0$ guarantees
that it is also affinely parametrized, which is the main reason for imposing this condition in the Kinnersley tetrad. 

The relevant equations in the NP formalism
are the Ricci and Bianchi identities written in terms of Weyl scalars and spin coefficients. 
Tetrad vectors can be gauged under the Lorentz group
of vector transformations. Given an algebraically general 
spacetime (Petrov type I), it is always possible to choose a tetrad where the two Weyl
scalars $\Psi_1$ and $\Psi_3$ vanish \cite{stephani2009exact}. This tetrad is not unique, and a detailed 
description of the properties of tetrads satisfying $\Psi_1=\Psi_3=0$ has been given in \cite{Nerozzi:2004wv}. In particular, it has been shown that
there are three infinite sets of transverse tetrads (transverse frames). $\Psi_0$ and $\Psi_4$
share the property of converging to zero in all of the tetrads constituting one the three
different transverse frames; for this reason this specific frame has been dubbed quasi-Kinnersley, because it must include the Kinnersley tetrad \cite{1969JMP....10.1195K} in the Petrov type D limit.

What makes each frame an infinite number of tetrads is the remaining 
choice of the spin/boost
parameter that leaves the condition $\Psi_1=\Psi_3=0$ unchanged.
A simple additional condition that removes this degeneracy is $\Psi_0=\Psi_4$. Assuming that the quasi-Kinnersley frame is 
considered, it is worth reminding that such a condition does 
not correspond to the Kinnersley tetrad in the type D limit, nevertheless it is an interesting
condition due to its simplicity, and we will adopt it for the calculations in this paper. 
This explains why we will be forced to reintroduce the spin/boost parameter in Sec.
\ref{sec:kerrexample} when we will compare our results for the spin coefficients with the 
already known values in the Kinnersley tetrad for a Kerr spacetime. 

Setting $\Psi_1=\Psi_3=0$ and $\Psi_0=\Psi_4$ completely fixes the tetrad up
to vector exchanges $\ell_a\leftrightarrow n_a$ and $m_a\leftrightarrow \bar{m}_a$
that leave these conditions unaltered. This additional freedom will not be 
removed in this work, but we will make sure that only variables that are not affected by it are
considered. 

Under such assumptions, the only remaining degrees of freedom in the Wey
scalars are 
$\Psi_2$ and $\Psi_4$. Their expression is given by $\Psi_2=-\frac{1}{2\sqrt{3}}\Psi_+$ and $\Psi_4=-\frac{i}{2}\Psi_-$, having defined

\begin{equation}
\Psi_{\pm}= I^{\frac{1}{2}}\left(e^{\frac{2\pi i k}{3}}\Theta\pm e^{-\frac{2\pi i k}{3}}\Theta^{-1}\right),
\label{eqn:pispm}
\end{equation}
and $\Theta=\sqrt{3}P I^{-\frac{1}{2}}$, $P=\left[-J+\sqrt{J^2-\left(I/3\right)^3}\right]^{\frac{1}{3}}$. $I$ and $J$ are the two curvature invariants defined as

\begin{subequations}
\label{eqn:curvinvariantsIJdef}
\begin{eqnarray}
I&=&\frac{1}{32}\:C^*_{abcd}\:C^{*abcd}, \label{eqn:curvinvariantsIWeyl}\\
J&=&\frac{1}{384}\:C^*_{abcd}\:{C^{*cd}}_{ef}\:C^{*abef}\label{eqn:curvinvariantsJWeyl},
\end{eqnarray}
\end{subequations}
and $k$ is an integer number 
that spans the interval $\left[0,1,2\right]$ identifying the three different transverse frames
(see \cite{Nerozzi:2004wv} for further details). In this study we consider the quasi-Kinnersley
frame, i.e. the only one in which $\Psi_-\rightarrow0$ in the Petrov type D limit.
In Eq.~(\ref{eqn:curvinvariantsIJdef}) $C^*_{abcd}$ is the self-dual form of the Weyl tensor
studied more in detail in section \ref{sec:NPselfdual}.

Eq.~(\ref{eqn:pispm}) shows that fixing completely the tetrad allows
to write the relevant remaining quantities as functions of tetrad
invariants. As the equation clearly states, this is true for the Weyl scalars. The work of this paper
aims at finding an analogous result for the spin coefficients. 

The curvature invariants $I$ and $J$ can be expressed in terms of the 
Weyl scalars, and within the specific tetrad choice considered here, they are given by

\begin{subequations}
\label{eqn:curvinvariantsIJ}
\begin{eqnarray}
I&=&\frac{1}{4}\left(\Psi_+^2-\Psi_-^2\right), \label{eqn:curvinvariantsI}\\
J&=&-\frac{\Psi_+}{24\sqrt{3}}\left(\Psi_+^2+3\Psi_-^2\right) \label{eqn:curvinvariantsJ}.
\end{eqnarray}
\end{subequations}

An alternative expression for the curvature invariant $J$ as a function of $I$ and $\Theta$ that will
be used in Sec. \ref{sec:dertetradindependent} is 

\begin{equation}
\label{eqn:jalternativeexpr}
J=-\frac{I^{\frac{3}{2}}}{6\sqrt{3}}\:\left(\Theta^3+\Theta^{-3}\right).
\end{equation}

Finally, it is useful to highlight an important function of curvature invariants given by
$\mathcal{S} = I^3-27J^2$. This function plays an important role in the study of the Petrov
type D limit as it tends to zero for a Kerr spacetime. Such a property becomes more evident 
when $\mathcal{S}$ is expressed as function of $I$ and $\Theta$, namely

\begin{equation}
\label{eqn:sinvarfuntheta}
\mathcal{S}=-\frac{I^3}{4}\:\left(\Theta^3-\Theta^{-3}\right)^2,
\end{equation}
and remembering that $\Theta\rightarrow 1$ in the Kerr limit.

\subsection{Bianchi identities}
\label{sec:bianchiidentitiessect}

With the choice of transverse tetrad adopted here, the Bianchi identities simplify to

\begin{subequations}
\label{eqn:bianchimod}
\begin{eqnarray}
D\Psi_{+}&=&i\sqrt{3}\:\lambda\:\Psi_{-}-3\:\rho\:\Psi_{+},
\label{eqn:bianchimodD4} \\
D\Psi_{-}&=&-i\sqrt{3}\:\lambda\:\Psi_{+}+\left(4\epsilon-\rho\right)\:\Psi_{-},
\label{eqn:bianchimodD2} \\
\Delta\Psi_{+}&=&-i\sqrt{3}\:\sigma\:\Psi_{-}+3\:\mu\:\Psi_{+},
\label{eqn:bianchimodDelta4} \\
\Delta\Psi_{-}&=&i\sqrt{3}\:\sigma\:\Psi_{+}-\left(4\gamma-\mu\right)\:\Psi_{-},
\label{eqn:bianchimodDelta2} \\
\delta\Psi_{+}&=&i\sqrt{3}\:\nu\:\Psi_{-}-3\:\tau\:\Psi_{+},
\label{eqn:bianchimoddelta4} \\
\delta\Psi_{-}&=&-i\sqrt{3}\:\nu\:\Psi_{+}+\left(4\beta-\tau\right)\:\Psi_{-},
\label{eqn:bianchimoddelta2} \\
\delta^*\Psi_{+}&=&-i\sqrt{3}\:\kappa\:\Psi_{-}+3\:\pi\:\Psi_{+},
\label{eqn:bianchimoddeltas4} \\
\delta^*\Psi_{-}&=&i\sqrt{3}\:\kappa\:\Psi_{+}-\left(4\alpha-\pi\right)\:\Psi_{-},
\label{eqn:bianchimoddeltas2} 
\end{eqnarray}
\end{subequations}

having defined the
directional derivatives $D = \ell^a\nabla_a$, $\Delta=n^a\nabla_a$, $\delta=m^a\nabla_a$ and
$\delta^*=\bar{m}^a\nabla_a$ along the tetrad vectors. 

Equation (\ref{eqn:bianchimod}) shows that the Bianchi identities can be considered as a linear system
to obtain the spin coefficients as functions of derivatives of the two tetrad invariants $\Psi_+$ and
$\Psi_-$. Such a system is however underdetermined as it consists of eight relations for twelve unknowns,
a result that was already found by Bonanos \cite{Bonanos:1996uy} in his paper on integrability of the NP equations. This raises the question whether it is possible to find 
other relations to close the system.
To answer this question, the NP formalism will be presented in the next section using
self-dual forms: this will lead to major simplifications in the formalism and allow an easier
characterisation of the missing relations. 

\section{Self-dual forms in the NP formalism}
\label{sec:NPselfdual}

\subsection{Self-dual forms and gravitational field}

As is well known (see for example \cite{stephani2009exact}) it is possible to introduce the following two-forms
as functions of the NP tetrad vectors 

\begin{subequations}
\label{eqn:twoformsdef}
\begin{eqnarray}
\Sigma_{ab}&=&2\:\ell_{[a}n_{b]}-2\:m_{[a}\bar{m}_{b]}, \\
\Sigma^{+}_{ab}&=&2\:\ell_{[a}m_{b]}, \\
\Sigma^-_{ab}&=&2\: n_{[a}\bar{m}_{b]}.
\end{eqnarray}
\end{subequations}

$\Sigma_{ab}$, $\Sigma^+_{ab}$ and $\Sigma^-_{ab}$ are self-dual, i.e. they satisfy the condition $\Sigma_{ab}=\frac{i}{2}\:{\epsilon_{ab}}^{cd}\:\Sigma_{cd}$,
$\epsilon_{abcd}$ being the Levi-Civita tensor, and can be thought as an alternative way of expressing the gravitational
field. The metric of the system is given by

\begin{equation}
g_{ab}=-\ell_an_b -n_a\ell_b+m_a\bar{m}_b+\bar{m}_am_b.
\end{equation}

Throughout 
this paper several calculations with contractions between $\Sigma_{ab}$,  $\Sigma^+_{ab}$ and
$\Sigma^-_{ab}$ 
will appear. Such contractions are just a consequence of the scalar products among NP tetrad vectors
and can be summarized by the following set of relations:

\begin{subequations}
\label{eqn:contractidentsigma}
\begin{eqnarray}
{\Sigma_a}^c\Sigma_{cb}&=&g_{ab},\\
{\Sigma_a}^c\Sigma^+_{cb}&=&-\Sigma^+_{ab},\\
{\Sigma_a}^c\Sigma^-_{cb}&=&\Sigma^-_{ab},\\
{\Sigma^+_a}^c\Sigma^+_{cb}&=&0,\\
{\Sigma^+_a}^c\Sigma^-_{cb}&=&\frac{1}{2}\left(g_{ab}-\Sigma_{ab}\right),\\
{\Sigma^-_a}^c\Sigma^-_{cb}&=&0.
\end{eqnarray}
\end{subequations}

In particular, if the remaining free indices are also contracted, the only nonvanishing relations are given by

\begin{subequations}
\label{eqn:contractidentsigma2indices}
\begin{eqnarray}
\Sigma^{ab}\Sigma_{ab}&=&-4,\\
\Sigma^{+ab}\Sigma^-_{ab}&=&-2.
\end{eqnarray}
\end{subequations}

Hereafter we will refer to $\Sigma_{ab}$,  $\Sigma^+_{ab}$ and
$\Sigma^-_{ab}$ as the gravitational field self-dual forms.

\subsection{Spin coefficients}
\label{sec:spincoeffsect}

The three groups of spin coefficients introduced in Sec. \ref{sec:bianchiidentitiessect} can be expressed
in a simplified way as projections of three fundamental vectors along the 
four tetrad vectors. To do so, the covariant
derivatives of the gravitational field self-dual forms introduced in the previous section
will be considered:

\begin{subequations}
\label{eqn:derivtwoforms}
\begin{eqnarray}
\nabla_{a}\Sigma_{bc}&=&2\cdot\left(T^+_{a}\:\Sigma^-_{bc}
-T^-_{a}\:\Sigma^+_{bc}\right) \\
\nabla_{a}\Sigma^+_{bc}&=&-T_{a}\:\Sigma^+_{bc}-
T^+_{a}\:\Sigma_{bc} ,\\
\nabla_{a}\Sigma^-_{bc}&=&T^-_{a}\:\Sigma_{bc} 
+T_{a}\:\Sigma^-_{bc},
\end{eqnarray}
\end{subequations}
where the vectors $T_a$, $T^+_{a}$ and $T^-_a$ are given by

\begin{subequations}
\label{eqn:tvectorsdef}
\begin{eqnarray}
T_{a}&=&n^{b}\:\nabla_{a}\ell_{b}+m^{b}\:\nabla_{a}\bar{m}_{b}, \\
T^+_{a}&=&\ell^{b}\:\nabla_{a}m_{b}, \\
T^-_{a}&=&n^{b}\:\nabla_{a}\bar{m}_{b}.
\end{eqnarray}
\end{subequations}

The vectors $T_a$, $T^+_a$ and $T^-_a$
constitute a compact way to express the NP spin coefficients, as
the latter can be derived projecting the former along the tetrad vectors, resulting
in twelve independent scalars as expected. However, the choice of $T_a$, $T^+_a$ and $T^-_a$
is not the most suitable one to write
them as functions of tetrad invariants, which is the main motivation
underlying this work. This is because, 
as already pointed out, the conditions 
$\Psi_1=\Psi_3=0$ and $\Psi_0=\Psi_4$ fix the tetrad up to the exchange operation
$\ell^a\leftrightarrow n^a$ and $m^a\leftrightarrow \bar{m}^a$. Unfortunately 
the vectors introduced in Eq.~(\ref{eqn:tvectorsdef}) are sensitive to the exchange operation
and transform according to 

\begin{subequations}
\label{eqn:tvectorsdeftranf}
\begin{eqnarray}
T_{a}&\rightarrow&-T_{a}, \\
T^+_{a}&\rightarrow&T^-_{a}, \\
T^-_{a}&\rightarrow&T^+_{a}.
\end{eqnarray}
\end{subequations}

Being sensitive to a tetrad change that does not affect the transverse conditions, they 
cannot be expressed as functions of tetrad
invariants. However, since the gravitational field self-dual forms transform under the same exchange
operation as

\begin{subequations}
\label{eqn:sigmavectorstransexchange}
\begin{eqnarray}
\Sigma_{ab}&\rightarrow&-\Sigma_{ab}, \\
\Sigma^+_{ab}&\rightarrow&\Sigma^-_{ab}, \\
\Sigma^-_{ab}&\rightarrow&\Sigma^+_{ab},
\end{eqnarray}
\end{subequations}
it is possible to construct a set of three derived vectors, namely

\begin{subequations}
\label{eqn:abcvectors}
\begin{eqnarray}
A_{a}&=&\Sigma^+_{ab}\:T^{-b}+\Sigma^-_{ab}\:T^{+b}, \label{eqn:abcvectorsa} \\
B_{a}&=&\Sigma^+_{ab}\:T^{+b}+\Sigma^-_{ab}\:T^{-b}, \\
C_{a}&=&\Sigma_{ab}\:T^{b},
\end{eqnarray}
\end{subequations}
which are now invariant under the exchange transformation, thus representing good candidates to be expressed
as functions of tetrad invariants.

The original NP spin coefficients are then given simply as

\begin{center}
$\begin{array}{ccc}
\rho=\ell^{a}A_{a}, & \lambda=-\ell^{a}B_{a}, &\epsilon=\frac{1}{2}\:\ell^{a}C_{a},  \\
 \mu=-n^{a}A_{a}, & \sigma=n^{a}B_{a},& \gamma=-\frac{1}{2}\:n^{a}C_{a}, \\
 \tau=m^{a}A_{a}, & \nu=-m^{a}B_{a},& \beta=\frac{1}{2}\:m^{a}C_{a},  \\
 \pi=-\bar{m}^{a}A_{a}, &\kappa=\bar{m}^{a}B_{a}, & \alpha=-\frac{1}{2}\:\bar{m}^{a}C_{a}.
\end{array}$
\end{center}

Hereafter we will refer to the three vectors $A_a$, $B_a$ and $C_a$ as connection vectors. 
With these definitions of the spin coefficients, the Bianchi identities given in Eq.~(\ref{eqn:bianchimod}) can be
rewritten in the compact form

\begin{subequations}
\label{eqn:bianchiidcompact}
\begin{eqnarray}
\nabla_a\Psi_+&=&-i\:\sqrt{3}\:\Psi_-\:B_a-3A_a\Psi_+, \label{eqn:bianchi2dentcvector} \\
\nabla_a\Psi_-&=&i\:\sqrt{3}\:\Psi_+\:B_a+\left(2C_a-A_a\right)\Psi_-. \label{eqn:bianchi1} 
\end{eqnarray}
\end{subequations}

\subsection{Quadratic self-dual forms and curvature}

We now turn to the curvature, and identify the relevant quantities for our study. To do so, 
a useful set quadratic self-dual tensors is introduced:

\begin{subequations}
\label{eqn:sigmatensorsnottrace}
\begin{eqnarray}
\Sigma_{abcd}&=&\Sigma_{ab}\:\:\Sigma_{cd}, \\
\Sigma^{++}_{abcd}&=&\Sigma^+_{ab}\:\:\Sigma^+_{cd}+\Sigma^-_{ab}\:\:\Sigma^-_{cd}, \\
\Sigma^{+-}_{abcd}&=&\Sigma^+_{ab}\:\:\Sigma^-_{cd}+\Sigma^-_{ab}\:\:\Sigma^+_{cd}.
\end{eqnarray}
\end{subequations}
Of the three tensors introduced in Eq.~(\ref{eqn:sigmatensorsnottrace}) one, namely $\Sigma^{++}_{abcd}$, is
trace-free, meaning that $g^{bd}\Sigma^{++}_{abcd}=0$. It is then possible to construct a linear combination
of the remaining two that is also trace-free:

\begin{equation}
\label{eqn:sigmatracefree}
\tilde{\Sigma}_{abcd}=\Sigma_{abcd}-\Sigma^{+-}_{abcd}.
\end{equation}

The tensors defined in Eqs.~(\ref{eqn:sigmatensorsnottrace}) and (\ref{eqn:sigmatracefree}) 
can be used as a basis to express relevant four rank tensors in this approach. 
The first tensor to be considered is the identity operator $I_{abcd}=\frac{1}{4}\left(g_{ac}g_{bd}-g_{ad}g_{bc}+i\epsilon_{abcd}\right)$
which is given in this basis by

\begin{equation}
I_{abcd}=-\frac{1}{4}\left(\Sigma_{abcd}+2\Sigma^{+-}_{abcd}\right).
\label{eqn:idoperatorsigmavar}
\end{equation}

The next step is to consider the curvature tensor. As only spacetimes in a vacuum 
are being considered here, the Weyl
tensor is the relevant quantity to define the curvature, its self-dual version being

\begin{equation}
\label{eqn:selfdualweylversion}
C^*_{abcd}=C_{abcd}+\frac{i}{2}  {\epsilon_{ab}}^{ef}C_{efcd}=2{I_{ab}}^{ef}C_{efcd}.
\end{equation}

The tensor $C^*_{abcd}$ can be projected along the basis of three self-dual
forms given in Eq.~(\ref{eqn:twoformsdef}), as shown e.g.
in \cite{stephani2009exact}. In transverse frames, 
where $\Psi_1=\Psi_3=0$ and
$\Psi_0=\Psi_4$, this leads to the following simple expression:

\begin{equation}
\label{eqn:weylform}
C^*_{abcd}=i\Psi_- \:\: \Sigma^{++}_{abcd}
+\frac{\Psi_+}{\sqrt{3}}\:\:\tilde{\Sigma}_{abcd}.
\end{equation}

For reasons that will be clearer in the following sections, 
it is also important to introduce a tensor
that has a quadratic dependence on the self-dual Weyl tensor. The most convenient 
choice for this purpose was found to be the Laplacian of the self-dual Weyl tensor defined as

\begin{equation}
\label{eqn:divselfdual}
D^*_{abcd}=\nabla_{\mu}\nabla^{\mu}C^*_{abcd}.
\end{equation}

In Sec. \ref{sec:bianchiselfdual} it will be shown that $D^*_{abcd}$ can be rewritten in an alternative way in which
the quadratic dependence on the self-dual Weyl tensor appears more evident.

The two tensors $C^*_{abcd}$ and $D^*_{abcd}$ share the same symmetries and are both trace-free.
They will be extensively used in the next sections. 

\section{Bianchi identities in the self-dual approach}
\label{sec:bianchiselfdual}

The Bianchi identities are given in vacuum by

\begin{equation}
\label{eqn:bianchicdual}
\nabla_{[a}C_{bc]de}=0
\end{equation}

Because of the symmetries of the Weyl tensor, Eq.~(\ref{eqn:bianchicdual}) holds also for its
self-dual version. The properties of self-dual tensors can be used to write
an alternative expression of Eq.~(\ref{eqn:bianchicdual}), namely

\begin{equation}
\label{eqn:bianchicdual2}
\nabla_{[a}C^*_{bc]de}=-\frac{i}{3}\:{\epsilon_{abc}}^f
\:\nabla_g {C^{*g}}_{fde}=0,
\end{equation}
so writing the Bianchi identities as $\nabla_{[a}C^*_{bc]de}=0$ or 
$\nabla_a{C^{*a}}_{bcd}=0$ is completely equivalent. 

It is possible to use the properties of the Weyl tensor described so far to find 
a useful alternative expression for the
tensor $D^*_{abcd}$ introduced in Eq.~(\ref{eqn:divselfdual}). This is achieved by writing

\begin{equation}
\label{eqn:dnewform}
D^*_{abcd}=2\:{I_{ab}}^{gh}\:{I_{cd}}^{il}\:\nabla_e\nabla_g{C^{*e}}_{hil},
\end{equation}
where the Bianchi identities on the indices $\{e,g,h\}$ have been enforced. 
Given that the term $\nabla_g\nabla_e{C^{*e}}_{hil}$ is vanishing thanks to the
Bianchi identities, one can replace the double covariant derivative in Eq.~(\ref{eqn:dnewform})
with its antisymmetrized version, yielding

\begin{equation}
\label{eqn:dnewform2}
D^*_{abcd}=4\:{I_{ab}}^{gh}\:{I_{cd}}^{il}\nabla_{[e}\nabla_{g]}{C^{*e}}_{hil}.
\end{equation}

Replacing the antisymmetrized derivative with the Weyl tensor gives

\begin{eqnarray}
\label{eqn:dnewform3}
D^*_{abcd}&=&{I_{ab}}^{gh}\:{I_{cd}}^{il}\:{C^*_{egh}}^f{C^{*e}}_{fil}\\&+&2\:{I_{ab}}^{gh}\:{I_{cd}}^{il}\:{C^*_{egi}}^f{C^{*e}}_{hfl}.\nonumber
\end{eqnarray}

There are several ways to simplify Eq.~(\ref{eqn:dnewform3}): one is to antisymmetrize in a suitable way the
indices of the Weyl tensors and then use the first type Bianchi identities. The other way is to perform
the calculation in transverse tetrads using Eq.~(\ref{eqn:weylform}) together with the contraction identities
in Eq.~(\ref{eqn:contractidentsigma}). Both ways lead to the final result

\begin{equation}
\label{eqn:dnewform5}
D^*_{abcd}=16 I \: I_{abcd}-\frac{3}{2}\:C^*_{abef}\:{C^{*ef}}_{cd},
\end{equation}
which shows explicitly the dependence of $D^*_{abcd}$ on the quadratic self-dual Weyl tensor.

Eq.~(\ref{eqn:dnewform5}) can be thought of as the self-dual version of the Penrose 
wave equation already introduced in \cite{1960AnPhy..10..171P} and originally given by

\begin{equation}
\label{eqn:penrosewaveeq}
\nabla_{\mu}\nabla^{\mu}C_{abcd}={C_{ab}}^{ef}\:C_{efcd}-4\:C_{aef[c}{{C^e}_{d]}}^{f}_{b}.
\end{equation}

 Several works have already analyzed interesting properties of this equation, in
 particular in \cite{1974PhRvD..10.1736R} it was shown that the Teukolsky equation can be derived
 from Eq.~(\ref{eqn:penrosewaveeq}). Here the tensor $D^*_{abcd}$ will be considered as a fundamental new variable.

In transverse frames, given Eqs.~(\ref{eqn:dnewform5}), (\ref{eqn:weylform}), (\ref{eqn:idoperatorsigmavar}) and the contraction identities 
in Eq.~(\ref{eqn:contractidentsigma}), the tensor $D^*_{abcd}$ takes the form

\begin{equation}
\label{eqn:dnewformtransverse}
D^*_{abcd}=-2i\sqrt{3}\Psi_+\Psi_-\:\: \Sigma^{++}_{abcd}+\left(\Psi^2_++\Psi_-^2\right)\: \tilde{\Sigma}_{abcd}.
\end{equation}

The tensor $D^*_{abcd}$ shares the
same symmetries with the Weyl tensor, and is also trace-free, 
allowing it to be expressed in the basis of the two trace-free tensors $\Sigma^{++}_{abcd}$ and $\tilde{\Sigma}_{abcd}$.

\section{Curvature and quadratic self-dual forms in transverse frames}
\label{sec:linkcurvselfdual}

In the previous two sections the self-dual form approach to the NP formalism has
been presented. In summary the following variables have been introduced to replace
the more familiar NP variables:

\begin{itemize}
\item The self-dual forms $\Sigma_{ab}$,  $\Sigma^+_{ab}$ and $\Sigma^-_{ab}$
as primary variables to characterise the gravitational field instead of the usual NP tetrad vectors.
\item The vectors $A_a$, $B_a$ and $C_a$ to identify the connection, having shown that the twelve spin
coefficients are the projections of these vectors along
the tetrad vectors. 
\item The self-dual Weyl tensor $C^*_{abcd}$ together with its 
Laplacian $D^*_{abcd}$ projected onto a suitable basis of quadratic self-dual forms.
\end{itemize}

Given the trace-free properties of  $C^*_{abcd}$ and $D^*_{abcd}$ it is useful to introduce
an alternative more convenient basis of quadratic trace-free self-dual forms given by the two tensors

\begin{subequations}
\label{eqn:sigmaupanddown}
\begin{eqnarray}
\Sigma^+_{abcd}&=&i\sqrt{3}\:\:\Sigma^{++}_{abcd}+\tilde{\Sigma}_{abcd}, \\
\Sigma^-_{abcd}&=&-i\sqrt{3}\:\:\Sigma^{++}_{abcd}+\tilde{\Sigma}_{abcd}.
\end{eqnarray}
\end{subequations}

The tensors $\Sigma^+_{abcd}$ and $\Sigma^-_{abcd}$ are just a linear
combinations of $\Sigma^{++}_{abcd}$ and $\tilde{\Sigma}_{abcd}$; it is 
therefore possible to use them as a basis for 
$C^*_{abcd}$ and $D^*_{abcd}$ 
using Eq.~(\ref{eqn:weylform})
and (\ref{eqn:dnewformtransverse}) (valid in transverse tetrads) 
together with the definition of $\Psi_+$ and
$\Psi_-$ given in Eq.~(\ref{eqn:pispm}). The result is given in matricial form by

\begin{equation}
\label{eqn:systemcdconv}
\begin{pmatrix} C^*_{abcd} \\ D^*_{abcd} \end{pmatrix}=
\textbf{U}
\begin{pmatrix} \Sigma^+_{abcd} \\ \Sigma^-_{abcd} \end{pmatrix},
\end{equation}

where

\begin{equation}
\label{eqn:matrthetasigma}
\textbf{U}=
\begin{pmatrix} 3^{-\frac{1}{2}}I^{\frac{1}{2}}\Theta& & 3^{-\frac{1}{2}}I^{\frac{1}{2}}\Theta^{-1} \\ 2I\Theta^{-2}& & 2I\Theta^{2}\end{pmatrix}.
\end{equation}

Equation (\ref{eqn:systemcdconv}) can of course be inverted to give the two tensors 
$\Sigma^+_{abcd}$ and $\Sigma^-_{abcd}$ as functions of $C^*_{abcd}$ and 
$D^*_{abcd}$, yielding 

\begin{equation}
\label{eqn:systemcdconvinv}
\begin{pmatrix} \Sigma^+_{abcd} \\ \Sigma^-_{abcd} \end{pmatrix}=
\textbf{U}^{-1}
\begin{pmatrix} C^*_{abcd} \\ D^*_{abcd} \end{pmatrix},
\end{equation}

where

\begin{equation}
\label{eqn:matrthetasigmainv}
\textbf{U}^{-1}=
\frac{\sqrt{3}}{2\:I^{\frac{3}{2}}\left(\Theta^3-\Theta^{-3}\right)}\begin{pmatrix} 2I\Theta^{2}& & -3^{-\frac{1}{2}}I^{\frac{1}{2}}\Theta^{-1} \\ -2I\Theta^{-2}& & 3^{-\frac{1}{2}}I^{\frac{1}{2}}\Theta\end{pmatrix}.
\end{equation}

As the primary motivation of this work is its application to numerical relativity, 
it is important to understand how the quantities introduced here behave in the Petrov
type D limit, in particular making sure that they are well defined. 
As already mentioned, the Kerr spacetime is obtained when 
$\Theta\rightarrow 1$. The expressions for $\Sigma^+_{abcd}$
and $\Sigma^-_{abcd}$ seem to be diverging in the limit if Eq.~(\ref{eqn:systemcdconvinv})
is taken into account. This is however
not the case as $C^*_{abcd}$ and $D^*_{abcd}$ cease to be independent 
in the limit. Using Eq.~(\ref{eqn:systemcdconv}) one can show that

\begin{equation}
D^*_{abcd}\left(\Theta\rightarrow 1\right)\rightarrow \:2\sqrt{3}\:I^{\frac{1}{2}}\:C^*_{abcd}.
\end{equation}

Given the degeneracy in the limit, it is important to understand what happens in its
neighbourhood by expanding in powers of $\left(\Theta-\Theta^{-1}\right)$ the tensor
$D^*_{abcd}$. This is done by writing

\begin{equation}
\label{eqn:dapproximated}
D^*_{abcd}\approx 2\sqrt{3}\:I^{\frac{1}{2}}\:C^*_{abcd}+\left(\Theta-\Theta^{-1}\right)\:{}^{(1)}D^*_{abcd},
\end{equation}
where

\begin{equation}
\label{eqn:dderivate}
{}^{(1)}D^*_{abcd}=\frac{\partial D^*_{abcd}}{\partial\left(\Theta-\Theta^{-1}\right)}=\lim_{\Theta\rightarrow 1}
\frac{D^*_{abcd}-2\sqrt{3}\:I^{\frac{1}{2}}\:C^*_{abcd}}{\Theta-\Theta^{-1}}.
\end{equation}

That the tensor ${}^{(1)}D^*_{abcd}$ is well defined can be proved by using the expressions
for $C^*_{abcd}$ and $D^*_{abcd}$ in transverse tetrads given by Eq.~(\ref{eqn:weylform}) and Eq.~(\ref{eqn:dnewformtransverse}). It is easy to show that

\begin{equation}
\lim_{\Theta\rightarrow 1}
\frac{D^*_{abcd}-2\sqrt{3}\:I^{\frac{1}{2}}\:C^*_{abcd}}{\Theta-\Theta^{-1}} = -6i\sqrt{3}\:I\:\Sigma^{++}_{abcd}.
\end{equation}
In other words, for $\Theta\rightarrow 1$, ${}^{(1)}D^*_{abcd}$ is proportional to the tensor $\Sigma^{++}_{abcd}$
in transverse tetrads, therefore it is well defined. 

Using Eq.~(\ref{eqn:systemcdconvinv}), (\ref{eqn:dapproximated}) and (\ref{eqn:dderivate}) 
we conclude that in the Kerr limit the two tensors $\Sigma^+_{abcd}$ and
$\Sigma^-_{abcd}$ are given by

\begin{subequations}
\label{eqn:sigmaplusminuslimit}
\begin{eqnarray}
\Sigma^+_{abcd}\left(\Theta\rightarrow 1\right)&\rightarrow&\frac{\sqrt{3}}{2\:I^{\frac{1}{2}}}\:C^*_{abcd}-\frac{1}{6\:I}\:{}^{(1)}D^*_{abcd}, \\
\Sigma^-_{abcd}\left(\Theta\rightarrow 1\right)&\rightarrow&\frac{\sqrt{3}}{2\:I^{\frac{1}{2}}}\:C^*_{abcd}+\frac{1}{6\:I}\:{}^{(1)}D^*_{abcd},
\end{eqnarray}
\end{subequations}
where all the diverging terms have been removed. 

\section{A suitable expression for the connection vectors}
\label{sec:tetradinvariantchar}

Having shown that the two tensor ${\Sigma^{+a}}_{bcd}$ and ${\Sigma^{-a}}_{bcd}$
constitute an optimal basis for the Weyl tensor and its Laplacian, it is important to relate
the connection vectors $A_a$, $B_a$ and $C_a$ to this basis. 
It is possible to do this by calculating the divergences
of ${\Sigma^{+a}}_{bcd}$ and ${\Sigma^{-a}}_{bcd}$ using the derivative identities
given in Eq.~(\ref{eqn:derivtwoforms}), yielding 

\begin{subequations}
\label{eqn:contractedsigmasigmaident}
\begin{eqnarray}
\nabla_a{\Sigma^{+a}}_{bcd}&=&2i\sqrt{3}T_a\left({\Sigma^{+a}}_b\Sigma^+_{cd}-{\Sigma^{-a}}_b\Sigma^-_{cd}\right) \\
&-&\left(i\sqrt{3}T^-_a+3T^+_a\right)\left({\Sigma^{-a}}_b\Sigma_{cd}+{\Sigma^{a}}_b\Sigma^-_{cd}\right)\nonumber  \\
&+&\left(i\sqrt{3}T^+_a+3T^-_a\right)\left({\Sigma^{+a}}_b\Sigma_{cd}+{\Sigma^{a}}_b\Sigma^+_{cd}\right), \nonumber \\
\nabla_a{\Sigma^{-a}}_{bcd}&=&-2i\sqrt{3}T_a\left({\Sigma^{+a}}_b\Sigma^+_{cd}-{\Sigma^{-a}}_b\Sigma^-_{cd}\right) \\
&+&\left(i\sqrt{3}T^-_a-3T^+_a\right)\left({\Sigma^{-a}}_b\Sigma_{cd}+{\Sigma^{a}}_b\Sigma^-_{cd}\right)\nonumber  \\
&-&\left(i\sqrt{3}T^+_a-3T^-_a\right)\left({\Sigma^{+a}}_b\Sigma_{cd}+{\Sigma^{a}}_b\Sigma^+_{cd}\right).\nonumber
\end{eqnarray}
\end{subequations}

Expressing the vectors $T_{a}$, $T^+_a$ and $T^-_a$ as functions of $A_a$, $B_a$ and
$C_a$ using Eq.~(\ref{eqn:abcvectors}) and simplifying using the contraction
identities given in Eq.~(\ref{eqn:contractidentsigma}) leads to the final result

\begin{equation}
\label{eqn:dsigmabartildematrix}
\nabla_a \begin{pmatrix} {\Sigma^{+a}}_{bcd} \\{\Sigma^{-a}}_{bcd} \end{pmatrix}=
\textbf{P}_a
\begin{pmatrix} {\Sigma^{+a}}_{bcd} \\{\Sigma^{-a}}_{bcd} \end{pmatrix},
\end{equation}
where

\begin{equation}
\label{eqn:coeffmatricsigmaplusminus}
\textbf{P}_a=\begin{pmatrix} 2A_a-C_a & A_a+C_a+i\sqrt{3}\:B_a  \\ A_a+C_a-i\sqrt{3}\:B_a & 2A_a-C_a  \end{pmatrix}.
\end{equation}

It is evident from Eq.~(\ref{eqn:dsigmabartildematrix}) that the divergences of the two tensors $\Sigma^+_{abcd}$
and $\Sigma^-_{abcd}$ identify uniquely the three connection vectors, as these can
be determined using the components $P^{11}_a$, $P^{12}_a$ and $P^{21}_a$
of the matrix $\textbf{P}_a$, namely

\begin{subequations}
\label{eqn:abcfuncp11}
\begin{eqnarray}
A_a&=& \frac{1}{6}\:\left(P_a^{21}+P_a^{12}+P_a^{11}\right), \\
B_a&=& \frac{i}{2\sqrt{3}}\:\left(P_a^{21}-P_a^{12}\right), \\
C_a&=& \frac{1}{3}\:\left(P_a^{21}+P_a^{12}-P_a^{11}\right).
\end{eqnarray}
\end{subequations}

As shown in Eq.~(\ref{eqn:systemcdconvinv}), it is possible to 
obtain the tensors $\Sigma^+_{abcd}$
and $\Sigma^-_{abcd}$ in transverse tetrads as a linear combination of tetrad invariant quantities like the self-dual Weyl 
tensor and its Laplacian. Having related the three connection vectors to the divergences
of $\Sigma^+_{abcd}$
and $\Sigma^-_{abcd}$ by means of Eqs.~(\ref{eqn:dsigmabartildematrix}) and (\ref{eqn:abcfuncp11}), in the next section we will combine the two results and derive
the connection vectors in transverse tetrads as a linear combination
of the divergences of $C^*_{abcd}$ and $D^*_{abcd}$. 

\section{Connection vectors in transverse tetrads}
\label{sec:dertetradindependent}

The linear relation between the quadratic self-dual basis given by 
$\Sigma^+_{abcd}$
and $\Sigma^-_{abcd}$ and the two tensors 
$C^*_{abcd}$ and $D^*_{abcd}$ given by Eq.~(\ref{eqn:systemcdconv}) can be used to obtain a suitable expression
for the divergences of the latter. Using 
Eqs.~(\ref{eqn:systemcdconv}), (\ref{eqn:dsigmabartildematrix}) and (\ref{eqn:systemcdconvinv}) 
this leads to the result written in matricial form as

\begin{equation}
\label{eqn:dsigmabartildematrixq}
\nabla_a \begin{pmatrix} {C^{*a}}_{bcd} \\{D^{*a}}_{bcd} \end{pmatrix}=
\textbf{Q}_a
\begin{pmatrix} {C^{*a}}_{bcd} \\{D^{*a}}_{bcd} \end{pmatrix},
\end{equation}
where

\begin{equation}
\label{eqn:ainvariantexpr}
\textbf{Q}_a=\nabla_a\left(\textbf{U}\right)\cdot\textbf{U}^{-1}+\textbf{U}
\cdot\textbf{P}_a\cdot\textbf{U}^{-1}.
\end{equation} 

Because of the Bianchi identities $\nabla_a{C^{*a}}_{bcd}=0$, the matrix $\textbf{Q}_a$
must take the form

\begin{equation}
\label{eqn:coeffmatricqts}
\textbf{Q}_a=\begin{pmatrix} 0 & 0  \\ \mathcal{S}_a & \mathcal{T}_a  \end{pmatrix},
\end{equation}
implying that the divergence of the tensor $D^*_{abcd}$ satisfies a relation 
of the type

\begin{equation}
\label{eqn:bianchityperelforD}
\nabla_a\:{D^{*a}}_{bcd}=\mathcal{S}_a\:{C^{*a}}_{bcd}+\mathcal{T}_a\:{D^{*a}}_{bcd},
\end{equation}
where the two vectors  $\mathcal{S}_a$ and $\mathcal{T}_a$ must be tetrad independent,
as they are relating tetrad invariants. It is possible to calculate 
their expression by contracting Eq.~(\ref{eqn:bianchityperelforD}) with the tensors
${C^*_{e}}^{bcd}$ and ${D^*_{e}}^{bcd}$ and using the contraction identities given in Eq.~(\ref{eqn:contractidentsigma}), yielding

\begin{subequations}
\label{eqn:systemst}
\begin{eqnarray}
C^*_{abcd}\:\nabla_e\:D^{*ebcd}&=&8\:I\:\mathcal{S}_a-144\:J\:\mathcal{T}_a, \label{eqn:systemst1}\\
D^*_{abcd}\:\nabla_e\:D^{*ebcd}&=&-144\:J\:\mathcal{S}_a+96\:I^2\:\mathcal{T}_a \label{eqn:systemst2}.
\end{eqnarray}
\end{subequations}

The term in the left hand side of Eq.~(\ref{eqn:systemst1}) 
is easily simplified by integrating by parts
and applying the Bianchi identities on $\nabla_e\:C^*_{abcd}$, the result being $-48\nabla_a\:J$.
The same trick cannot be used for the left hand side of Eq.~(\ref{eqn:systemst2}). Upon defining 

\begin{equation}
\mathcal{R}_a=\frac{1}{96}\:D^*_{abcd}\:\nabla_e\:D^{*ebcd},
\end{equation} 
the system in Eq.~(\ref{eqn:systemst}) can be inverted to give

\begin{subequations}
\label{eqn:systemstbisinv}
\begin{eqnarray}
\mathcal{S}_a&=&\frac{1}{S}\left(-6\:I^2\:\nabla_a\:J+18\:J\:\mathcal{R}_a\right), \label{eqn:systemst1bisinv}\\
\mathcal{T}_a&=&\frac{1}{S}\left(-9\:J\:\nabla_aJ+I\:\mathcal{R}_a\right) \label{eqn:systemst2bisinv},
\end{eqnarray}
\end{subequations}
where $S$ is the scalar curvature invariant defined in Eq.~(\ref{eqn:sinvarfuntheta}). 
Given that $S\rightarrow0$ in the Petrov type D limit, 
Eq.~(\ref{eqn:systemstbisinv}) may appear to diverge in this case.
However, analogously to what was done in section \ref{sec:linkcurvselfdual} for $\Sigma^+_{abcd}$ and $\Sigma^-_{abcd}$, it is
possible to verify that $\mathcal{S}_a$ and $\mathcal{T}_a$ are well defined in the Kerr limit.  

Equation (\ref{eqn:systemst1}) can be rewritten in the alternative form 

\begin{equation}
\label{eqn:tssimple}
\mathcal{T}_a=-\tilde{S}_a+\nabla_a\:\ln\left[I^{\frac{1}{2}}\left(\Theta^3+\Theta^{-3}\right)^{\frac{1}{3}}\right],
\end{equation}
where the expression for the curvature invariant $J$ given
in Eq.~(\ref{eqn:jalternativeexpr}) has been used together with the identity

\begin{equation}
\label{eqn:resuceds}
\tilde{S}_a=\frac{I^{-\frac{1}{2}}}{\sqrt{3}\left(\Theta^3+\Theta^{-3}\right)}\:\mathcal{S}_a.
\end{equation}

Assuming that the vector $\mathcal{S}_a$ is well defined in the Petrov type D limit,
Eq.~(\ref{eqn:tssimple}) ensures that $\mathcal{T}_a$ is also well defined 
when $\Theta\rightarrow 1$, meaning that it 
is only necessary to check the behavior of  
$\mathcal{S}_a$. For this purpose Eq.~(\ref{eqn:systemst1bisinv}) has to be analyzed more in detail, and
in particular the expression for $\mathcal{R}_a$. Given that the scalar $S$ in the denominator
of Eq.~(\ref{eqn:systemst1bisinv}) has a singular term $\frac{1}{\left(\Theta-\Theta^{-1}\right)^2}$
it is important that no terms of zero and first order in $\left(\Theta-\Theta^{-1}\right)$ appear
in the numerator. Expanding $\mathcal{R}_a$ as

\begin{equation}
\mathcal{R}_a\approx\mathcal{R}^{(0)}_a+\left(\Theta-\Theta^{-1}\right)\mathcal{R}^{(1)}_a+\left(\Theta-\Theta^{-1}\right)^2\mathcal{R}^{(2)}_a,
\end{equation}
and considering Eq.~(\ref{eqn:dapproximated}) to express 
the tensor $D^*_{abcd}$ in powers of $\left(\Theta-\Theta^{-1}\right)$, it is possible
to obtain the terms $\mathcal{R}^{(0)}_a$ and $\mathcal{R}^{(1)}_a$ given by

\begin{subequations}
\label{eqn:rexpsnd}
\begin{eqnarray}
\mathcal{R}^{(0)}_a&=&\frac{1}{2}\:I\:\nabla_a I, \\
\mathcal{R}^{(1)}_a&=&0.
\end{eqnarray}
\end{subequations}

The term $\mathcal{R}^{(0)}_a$ eliminates the other zero order term in Eq.~(\ref{eqn:systemst1bisinv}), keeping in mind that in the Kerr limit $J=-\frac{I^{\frac{3}{2}}}{3\sqrt{3}}$,
as obtained from Eq.~(\ref{eqn:jalternativeexpr}). We conclude that only terms of power $\left(\Theta-\Theta^{-1}\right)^2$
appear in the numerator of Eq.~(\ref{eqn:systemst1bisinv}), thus ensuring that $\mathcal{S}_a$
is well defined in the Petrov type D limit.  
 
Having found that $ \mathcal{S}_a$ and $\mathcal{T}_a$ can be given as functions
of tetrad invariants, and having verified that these functions are well behaved in the Petrov
type D limit, it is now possible to obtain an alternative expression for the matrix $\textbf{P}_a$
defined in Eq.~(\ref{eqn:dsigmabartildematrix}) only using tetrad invariants. The procedure is identical to the one adopted to obtain Eq.~(\ref{eqn:dsigmabartildematrixq}), just applied in the opposite direction: Eq.~(\ref{eqn:systemcdconvinv}) is the starting point to express $\Sigma^+_{abcd}$ and $\Sigma^-_{abcd}$
as functions of $C^*_{abcd}$ and $D^*_{abcd}$, then Eq.~(\ref{eqn:dsigmabartildematrixq}) to eliminate the divergences
of the Weyl tensor and its Laplacian, and finally Eq.~(\ref{eqn:systemcdconv}) to rewrite everything
in function of $\Sigma^+_{abcd}$ and $\Sigma^-_{abcd}$. The result is

\begin{equation}
\label{eqn:ainvariantexprpa}
\textbf{P}_a=\nabla_a\left(\textbf{U}^{-1}\right)\cdot\textbf{U}+\textbf{U}^{-1}
\cdot\textbf{Q}_a\cdot\textbf{U}.
\end{equation} 

Written explicitly in components, the matrix $\textbf{P}_a$ is given by

\begin{equation}
\label{eqn:coeffmatricqtspa}
\textbf{P}_a=\begin{pmatrix}\textbf{P}^{11}_a & \textbf{P}^{12}_a  \\ \textbf{P}^{21}_a & \textbf{P}^{22}_a  \end{pmatrix},
\end{equation}
where

\begin{subequations}
\label{eqn:compomentsptutti}
\begin{eqnarray}
\textbf{P}^{11}_a&=&-\frac{1}{2}\:\left[\tilde{S}_a+\nabla_a\:\ln\:\left(I\:K\right)\right], \\
\textbf{P}^{12}_a&=&\frac{\Theta^{-2}}{2}\:\left[\tilde{S}_a+\nabla_a\ln\:\left(\Theta^2\:K\right)\right], \\
\textbf{P}^{21}_a&=&\frac{\Theta^{2}}{2}\:\left[\tilde{S}_a+\nabla_a\ln\:\left(\Theta^{-2}\:K\right)\right], \\
\textbf{P}^{22}_a&=&-\frac{1}{2}\left[\tilde{S}_a+\nabla_a\:\ln\:\left(I\:K\right)\right],
\end{eqnarray}
\end{subequations}
and

\begin{equation}
K=\frac{\Theta^3-\Theta^{-3}}{\left(\Theta^3+\Theta^{-3}\right)^{\frac{1}{3}}}. \label{eqn:definitK} 
\end{equation}

As expected from Eq.~(\ref{eqn:coeffmatricsigmaplusminus}), the two components 
$\textbf{P}^{11}_a$ and $\textbf{P}^{22}_a$ coincide.

Putting together Eqs.~(\ref{eqn:abcfuncp11}), (\ref{eqn:coeffmatricqtspa}) and (\ref{eqn:compomentsptutti}) gives the final expression for 
the connection vectors in transverse tetrads:

\begin{subequations}
\label{eqn:finalABC}
\begin{eqnarray}
A_a&=&\frac{\mathcal{E}_A}{12}\left[\tilde{S}_a+\nabla_a\ln\left(\frac{K}{\mathcal{E}_A}\right)\right]-\frac{1}{6}\nabla_a\ln I,  \\
B_a&=&\frac{i\:\mathcal{E}_B}{4\sqrt{3}}\:\left[\tilde{S}_a+\nabla_a\ln\left(\frac{K}{\mathcal{E}_B}\right)\right], \\
C_a&=&\frac{\mathcal{E}_C}{6}\left[\tilde{S}_a+\nabla_a\ln\left(\frac{K}{\mathcal{E}_C}\right)\right]+\frac{1}{6}\nabla_a\ln I.
\end{eqnarray}
\end{subequations}
where $\mathcal{E}_A=\left(\Theta-\Theta^{-1}\right)^2$, $\mathcal{E}_B=\Theta^2-\Theta^{-2}$ and $\mathcal{E}_C=\Theta^2+\Theta^{-2}+1$. 

This completes the demonstration and shows 
that it is possible to 
fix all the spin coefficients in the NP formalism once the tetrad is unambiguously chosen. 
The additional information on the divergence of a 
quadratic function of the self-dual Weyl tensor was found to be crucial to solve the system. 
It was shown that such a quantity naturally introduces 
a third tetrad invariant vector ($\mathcal{S}_a$) that is independent of the 
derivatives of the two curvature invariants $I$ and $J$. A more detailed study of the properties
of $\mathcal{S}_a$ and of Eq.~(\ref{eqn:bianchityperelforD}) using a coordinate based approach
will be given in a follow-up paper.

\section{The Kerr limit}
\label{sec:kerrexample}

An important aspect of our study is to verify how the results found behave in the single
black hole limit. The value of the spin coefficients in the Kerr spacetime 
is well known using the Kinnersley tetrad. As already mentioned 
in section \ref{eqn:weylscalarscurvatureinv}, we expect a spin/boost transformation between the tetrad studied in this work and
the Kinnersley tetrad in the Petrov type D limit. Fortunately,  as will be shown here, 
only the connection vector $C_a$ is affected by this additional 
spin/boost transformation. 

Using Eq.~(\ref{eqn:finalABC}) it is easy to show that for $\Theta\rightarrow 1$ the connection vectors
are given by

\begin{subequations}
\label{eqn:finalABCtypeD}
\begin{eqnarray}
A_a&=&-\frac{1}{6}\:\nabla_a\ln\:I\label{eqn:finalABCcaseA}\\
B_a&=&0, \label{eqn:finalABCcaseB} \\
C_a&=&\frac{1}{6}\:\nabla_a\ln\:I + Z_a, \label{eqn:finalABCcaseC}
\end{eqnarray}
\end{subequations}
where $Z_a=\frac{1}{2}\left(\tilde{S}_a+\nabla_a\ln\:K\right)$ in the Petrov type D limit.
It is worth noticing that the vector $\nabla_a\ln\:K$ is equal to $\nabla_a\ln\left(\Theta-\Theta^{-1}\right)$ in the limit, therefore undefined at first sight. 
That this term is indeed well defined and not diverging in the Kerr limit is proved 
by the value of all the spin coefficients already known in the Kinnersley tetrad,
having already shown that $\mathcal{S}_a$ is also well defined. This is however a
point that requires further understanding. By means of the equations studied so far, namely the Bianchi
identities and Eq.~(\ref{eqn:bianchityperelforD}), it is not possible to gain more
information on this term, especially for what concerns its Petrov type D limit. However, other equations have to be considered within this simplified
approach to have a complete picture, like for example the Ricci identities. We expect that a full
understanding of all the equations that play a relevant role in the formalism will help
clarify this specific point too. This is the subject of future work. 

Some known results
follow in a straightforward manner from Eq.~(\ref{eqn:finalABCtypeD}). For example, the Goldberg-Sachs
theorem \cite{Goldberg:2009hv} (or, more precisely, a corollary of it applied to type D spacetimes) is
summarized by Eq.~(\ref{eqn:finalABCcaseB}), implying that the spin 
coefficients 
$\lambda$,
$\sigma$, $\nu$ and $\kappa$ vanish in the limit, which is exactly what the
theorem states. Any spin/boost transformation does not alter this result which must
continue to hold in the Kinnersley tetrad as expected.

The next step is to verify Eq.~(\ref{eqn:finalABCcaseA}) and the corresponding spin coefficients. To do so,
we consider the explicit expression of the metric of a Kerr spacetime using Boyer-Lindquist coordinates, i.e.

\begin{eqnarray}
ds^2&=&-\left(1-\frac{2Mr}{\Sigma}\right)dt^2+\left(\frac{4Mar\sin^2\theta}
{\Sigma}\right)dtd\phi+\left(\frac{\Sigma}{\Gamma}\right)dr^2 \nonumber \\
&+&\Sigma d\theta^2+\sin^2\theta\left(\frac{r^2+a^2+2Mar\sin^2\theta}{\Sigma}
\right)d\phi^2,
\label{eqn:kerrboyer}
\end{eqnarray}
where $\Gamma=r^2-2Mr+a^2$ 
(in the usual notation this quantity is referred to as $\Delta$, but here we
changed notation to avoid confusion with the derivative operator $\Delta$),
$\Sigma=r^2+a^2\cos^2\theta$, $M$ is the black hole mass and $a$ its
rotation parameter.

The expression for the Kinnersley tetrad vectors is

\begin{subequations}
\label{eqn:kerrtetrad}
\begin{eqnarray}
\ell^{\mu}&=&\left[-\left(r^2+a^2\right)/\Gamma,1,0,a/\Gamma\right], 
\label{eqn:kerrtetradl} \\
n^{\mu}&=&\left[-r^2-a^2,-\Gamma,0,a\right]/\left(2\Sigma\right), 
\label{eqn:kerrtetradn} \\
m^{\mu}&=&\left[-ia\sin\theta,0,1,i/\sin\theta\right]\cdot \rho^*/\sqrt{2},
\label{eqn:kerrtetradm}
\end{eqnarray}
\end{subequations}
where $\rho^*=\frac{1}{r+ia\cos\theta}$.

The curvature invariant $I$ is given by

\begin{equation}
\label{eqn:limitI}
I = \frac{3M^2}{\left(r-ia\cos\theta\right)^6}.
\end{equation}

Given Eqs.~(\ref{eqn:finalABCcaseA}) and (\ref{eqn:limitI}) the components of the
connection vector $A_a$ are given by

\begin{equation}
\label{eqn:alimitI}
A_a = \left[0, \frac{1}{r-ia\cos\theta},\frac{ia\sin\theta}{r-ia\cos\theta},0\right].
\end{equation}

It is possible to show from Eq.~(\ref{eqn:abcvectorsa}) that a spin/boost transformation does not affect
the vector $A_a$, therefore 
the projection of $A_a$ 
along the four null vectors must give the spin coefficients $\rho$, $\mu$, $\tau$ and $\pi$ 
as obtained in the Kinnersley tetrad. A simple calculation yields

\begin{center}
$\begin{array}{ccc}
\rho = \frac{1}{r-ia\cos\theta}, & &\mu = \frac{\rho\Gamma}{2\Sigma},  \\
\tau =\frac{ ia\rho\rho^*\sin\theta}{\sqrt{2}},& & \pi = -\frac{ia\rho^2\sin\theta}{\sqrt{2}}.
\end{array}$
\end{center}
in agreement with the known values in the Kinnersley tetrad. 

The final calculation for the vector $C_a$ is slightly more complicated
as this is the only vector that is affected by spin/boost transformations. A spin/boost rotation with complex parameter $\mathcal{B}$ affects the spin coefficients associated
to $C_a$ as

\begin{subequations}
\label{eqn:spinkintetr}
\begin{eqnarray}
\epsilon &=& \frac{1}{2}\:\ell^a\:\left(C_a-\nabla_a\ln\mathcal{B}\right), \\
\gamma&=&-\frac{1}{2}\:n^a\:\left(C_a+\nabla_a\ln\mathcal{B}\right), \label{eqn:spinkintetrgamma}\\
\beta &=& \frac{1}{2}\:m^a\:\left(C_a-\nabla_a\ln\mathcal{B}\right), \\
\alpha&=&-\frac{1}{2}\:\bar{m}^a\:\left(C_a+\nabla_a\ln\mathcal{B}\right) \label{eqn:spinkintetralpha}.
\end{eqnarray}
\end{subequations}

The Kinnersley tetrad corresponds to $\epsilon=0$, giving
the condition for $\mathcal{B}$,

\begin{equation}
\label{eqn:condbalongell}
\ell^a\:\nabla_a\ln\mathcal{B}= \ell^a \:C_a
\end{equation}

This condition only fixes the radial components of the gradient of $\mathcal{B}$ as
the vector $\ell^a$ only has nonvanishing components along the $t$ and $r$ direction, the $t$ 
direction giving no contribution because of the stationarity of the spacetime. If we considered
Eq.~(\ref{eqn:condbalongell}) as a general identity valid along all null vectors, i.e 
$\nabla_a\ln\mathcal{B}= C_a$, we would end
up with the spin coefficient $\beta=0$. This is not the case, given that the expression
for $\beta$ in the Kinnersley tetrad is given by

\begin{equation}
\label{eqn:expressionbetakin}
\beta = -\frac{\rho^*\:\cot\theta}{2\sqrt{2}}.
\end{equation}

Equation (\ref{eqn:expressionbetakin}) is a clear sign that there is an additional contribution
to $\nabla_a\ln\mathcal{B}$ along the $\theta$ direction. It is very easy to calculate this
additional contribution from the expression of $\beta$, leading to the final result,

\begin{equation}
\label{eqn:spinvaluefinal}
\nabla_a\ln\mathcal{B}=C_a+\nabla_a \ln\:\sin\theta.
\end{equation}

Having determined the spin/boost parameter, we can substitute it into the remaining 
Eqs.~(\ref{eqn:spinkintetrgamma}) and (\ref{eqn:spinkintetralpha}) to find their
expression in the Kinnersley tetrad using Eq.~(\ref{eqn:finalABCcaseC}), yielding

\begin{subequations}
\label{eqn:derivedgammaalpha}
\begin{eqnarray}
\gamma &=& -\mu  -n^a\:Z_a, \\
\alpha&=&-\pi  + \beta^* -\bar{m}^a\:Z_a.
\end{eqnarray}
\end{subequations}

Equation (\ref{eqn:derivedgammaalpha}) must be compared with 
the values for $\gamma$ and $\alpha$ in the Kinnersley tetrad given by

\begin{subequations}
\label{eqn:effectivevaluealphagamma}
\begin{eqnarray}
\gamma &=&\mu+\frac{1}{2}\:n^a\nabla_a\ln\Gamma, \\
\alpha &=& \pi-\beta^*.
\end{eqnarray}  
\end{subequations}

We find that Eq.~(\ref{eqn:effectivevaluealphagamma}) is compatible with Eq.~(\ref{eqn:derivedgammaalpha}) if the vector $Z_a$ is given by  

\begin{equation}
\label{eqn:svectorfinval}
Z_a = -\nabla_a\ln\left(\Gamma^{\frac{1}{2}}\:I^{\frac{1}{3}}\:\sin\theta\right).
\end{equation}

Combining Eq.~(\ref{eqn:svectorfinval}) with Eq.~(\ref{eqn:finalABCcaseC}) gives the actual value of the $C_a$ vector in the Kerr limit:

\begin{equation}
\label{eqn:cvectorfinval}
C_a = -\nabla_a\ln\left(\Gamma^{\frac{1}{2}}\:I^{\frac{1}{6}}\:\sin\theta\right),
\end{equation}
while the spin/boost parameter $\mathcal{B}$, using Eq.~(\ref{eqn:spinvaluefinal}), is given by

\begin{equation}
\label{eqn:finvalspiboo}
\mathcal{B} = \mathcal{B}_0\:I^{-\frac{1}{6}}\:\Gamma^{-\frac{1}{2}},
\end{equation}
with $\mathcal{B}_0$ being an integration constant. 
Equation (\ref{eqn:finvalspiboo}) is in agreement with the result already found in \cite{Nerozzi:2008ng} using a slightly different approach. With a well defined expression 
for the vector $S_a$ in a general Petrov type I spacetime, one could have enforced the
condition $\epsilon=0$ to obtain the spin/boost parameter between the tetrad considered 
in this paper and the quasi-Kinnersley tetrad. Lacking such an expression, we were only able to
obtain 
$S_a$ by comparing the values of the spin coefficients in the Petrov type D limit. 
As a consequence of this, the spin/boost parameter that was found depends on functions
that are only defined in the limit, like $\Gamma$, 
whose numerical implementation is complicated, because they are defined in a
specific coordinate system. A follow-up paper will fill this gap and give a general
expression for $S_a$ by studying Eq.~(\ref{eqn:bianchityperelforD}) in detail using a
standard coordinate based approach.


\section{Conclusions}
\label{sec:concl}

Many years after its introduction, the Newman-Penrose formalism continues to be
employed in many applications of Einstein's equations, and its use is
widely spread in the fields of theoretical and numerical relativity. A certain number of questions are 
nevertheless still open, 
in particular the possibility to simplify the formalism by removing all of the gauge degrees of freedom
and express all of the remaining quantities as functions of tetrad invariants. 
This is certainly possible for the Weyl scalars as pointed out in
\cite{Bonanos:1991dl} and shown in Eq.~(\ref{eqn:pispm}) of this paper, but no equivalent
result for the spin coefficients was known, although a similar argument must hold.  

Motivated by this, the aim of this paper was to prove that it is indeed possible to fix 
all of the spin coefficients as functions of tetrad invariants once the gauge freedom has
been completely removed. 
While previous works
had already shown that the Bianchi identities could fix eight of the twelve spin coefficients, 
the question on how to fix the remaining four remained unanswered. 
Here it was found that
the divergence of the Laplacian of the Weyl tensor, or more generally of a quadratic function of 
the Weyl tensor, is crucial to give the missing information, as it must satisfy a relation of the type

\begin{equation}
\label{eqn:bianchityperelforDbis}
\nabla_a\:{D^{*a}}_{bcd}=\mathcal{S}_a\:{C^{*a}}_{bcd}+\mathcal{T}_a\:{D^{*a}}_{bcd},
\end{equation}
which uniquely identifies a third tetrad invariant vector ($\mathcal{S}_a$) that cannot be obtained from the derivatives
of the two curvature invariants $I$ and $J$. While it was possible to identify this additional
vector and relate the spin coefficients to it, a general expression for $\mathcal{S}_a$ is still
lacking, and in particular, it is not yet known whether $\mathcal{S}_a$ can be expressed
as the gradient of a third tetrad invariant scalar function. 

As Eq.~(\ref{eqn:bianchityperelforDbis}) relates tetrad invariant quantities,
it can be obtained using standard coordinate based approaches. We will explore
this alternative approach in a forthcoming paper, aiming to derive 
a general expression for the vector $\mathcal{S}_a$. 

So how is this all relevant to numerical relativity? The answer is simple: 
with a well defined expression for $\mathcal{S}_a$ in a tetrad with $\Psi_1=\Psi_3=0$ and
$\Psi_0=\Psi_4$, it is possible to enforce the condition $\epsilon=0$ to find the spin/boost
parameter $\mathcal{B}$ between this tetrad and the quasi-Kinnersley tetrad. This last
information will allow us 
to write $\Psi_4$ in the right quasi-Kinnersley
tetrad as

\begin{equation}
\Psi_4^{QKT}=-\frac{i\:\mathcal{B}^2}{2}\: \Psi_-.
\end{equation}

In other words it will be possible to have a scalar quantity written as function of tetrad
invariants and defined in a general Petrov type I spacetime that naturally converges to the specific $\Psi_4$ studied in the perturbative
regime (Teukolsky equation) when the spacetime converges to Kerr, making it an fundamental 
gauge invariant quantity for numerical relativity and gravitational wave extraction. 

Besides the numerical applications which constitute the main motivation
of this work, it is stressed here that this methodology has great potentialities for a deeper
understanding of tetrad approaches to Einstein's equations, as
the number of relevant variables is reduced drastically. For this reason, 
having already given a simplified expression for the Bianchi
identities [Eq.~(\ref{eqn:bianchiidcompact})], future work will focus
on studying the properties of the Ricci identities within this same approach, thereby completing
the picture of relevant equations.


\acknowledgments

The author has been funded by the Funda\c c\~ao para a Ci\^encia e Tecnologia 
through Grant No. SFRH/BPD/103594/2014. This project has received funding from the European Union's
Horizon 2020 research and innovation programme under
the Marie Sklodowska-Curie grant agreement No 690904.



\bibliographystyle{apsrev}
\bibliography{/Users/andrea/Dropbox/Documents/Tex/Papers/SpinCoeffNP/SpinCoeffNP_fifth_version/references}


\end{document}